# On possible reconciliation of anomalous isotope effect and non-phononic pairing mechanism in high-$T_c$ cuprates

S. H. Naqib* and R. S. Islam

*Department of Physics, University of Rajshahi, Rajshahi-6205, Bangladesh*

Superconductivity in high-$T_c$ cuprates is unconventional and there is a growing acceptance that carrier pairing in these strongly correlated electronic materials is mediated mainly by bosons of electronic origin rather than the phonons. Still, a significant isotope exponent (IE) (in certain cases, much larger than 0.5, the canonical BCS value) is observed experimentally. This has led to the assumption that phonons are also involved up to a certain extent. The magnitude of the IE in cuprates depends strongly on the number of doped holes, $p$, in the $CuO_2$ planes and therefore, may indicate that phonons play different quantitative roles as the mediating boson at different regions of the $T_c$-$p$ phase diagram. Based on recent experiments on Fermi-surface reconstruction from the thermoelectric transport measurements (Nature Commun. **2:432** doi: 10.1038/ncomms1440 (2011)) and on magnetic field induced charge order in $YBa_2Cu_3O_{7-\delta}$ (Nature **477**, 191 (2011)), we propose a very simple scenario where isotope substitutions affect the *stripe state* via the coupling to the lattice and thereby change the superconducting $T_c$. In this picture significant part of the observed IE, over an extended $p$-range, originates from the isotope induced stripe modulation and is not directly related to the characteristic energy scale of the pairing phonons as is the case in the BCS framework. This simple proposal qualitatively explains all the $p$-dependent anomalous features of the IE observed experimentally in hole doped high-$T_c$ cuprates.

The pairing mechanism leading to the emergence of high-$T_c$ superconductivity in cuprates remains as one of the outstanding problems in the field of strongly correlated electronic systems. It is widely accepted that the normal and superconducting (SC) state properties of these materials are non-Fermi liquid like [1, 2] and the conventional BCS formalism [3] does not describe all the features shown at and below the SC transition temperature. The isotope effect, which played a defining role in the development of the BCS theory within the framework of electron-phonon interaction, is highly confusing in the case of cuprate superconductors [4 – 7]. The role played by phonons as the mediating bosons for Cooper pairing in cuprates is a matter of strong debate [5 – 9]. In these high-$T_c$ materials, the value of the isotope exponent (IE), $\alpha$, depends strongly on the number of added holes in the $CuO_2$ planes, $p$. The general trend being, $\alpha(p)$ is large inside a certain doping range in the underdoped (UD) side and can greatly exceed the canonical BCS value of 0.5 near the so-called $1/8^{th}$ anomaly [5, 6, 10], the value of IE near the optimum doping, on the other hand, is quite small and stays almost the same in the overdoped (OD) region [5, 6, 10]. The question that naturally arises is that whether a larger/smaller value of the IE corresponds to a greater/lesser contribution of the phonons in the pairing mechanism? No definitive answer exists so far.

Interpretation of the IE is highly complicated because of the various electronic correlations present in hole doped cuprates. The IE is affected by the pseudogap (PG), spin/charge ordering (stripe correlations), and possibly by other exotic density waves in the quasi-particle (QP) energy spectrum [10 – 12]. There exists a number of theoretical schemes aimed to describe the doping and disorder dependence of the IE in hole doped cuprates [5, 7, 10] but none of these formulations explain the doping and disorder dependent IE over the entire range of hole contents [4]. Since the early days of the discovery of high-$T_c$ cuprates, there has been a growing prominence of pairing theories based on spin fluctuations [13 – 16]. Recent optical spectroscopic study [17] on Bi2212 crystals have revealed the importance of the role of non-phononic electronic excitations (antiferromagnetic spin fluctuations) in Cooper pairing. How one reconciles the apparently contradictory experimentally observed significant IE with the findings from the optical study [17] poses a serious problem. In this short communication we wish to address this issue by taking into consideration solely the stripe correlations.

Besides superconductivity and the PG correlations, the spin/charge ordered *stripe state* is probably the most widely studied phenomenon in the field of research on cuprates. The static spin/charge stripe correlations are found in UD cuprates in the vicinity of $p \sim 0.125$ (the so-called $1/8^{th}$ anomaly) [18], although dynamical (fluctuating) stripe correlations are believed to exist over a much wider doping range, especially in the single $CuO_2$ layer LSCO compounds [11, 19]. Incommensurate low-energy spin fluctuations, generally interpreted as precursor to strong stripe correlations, are also observed in double-layer Y123 and Bi2212 [19, 20]. Therefore, there are reasons to believe that stripe ordering is a generic feature of hole doped cuprate superconductors. Spin/charge stripe probably forms in doped Mott insulators as a compromise between the AFM ordering among the Cu spins and strong Coulomb repulsion between the electrons (both favoring localization) with the kinetic energy of the mobile doped carriers (leading to delocalization). Broadly speaking, stripe phase can be viewed as spontaneously separated ordered states of charge-rich (high kinetic energy) and charge-poor (antiferromagnetically correlated) regions throughout the compound. It is interesting to note that the stripe dynamics is closely linked with the underlying lattice structure and any modifications/distortions in lattice dynamics plays a significant role in the pinning strength of fluctuating stripe order [21]. In other words, in the stripe phase the lattice and the charge/spin degrees of freedoms are strongly intertwined [19, 21, 22]. The possible link between stripe and SC correlations are not entirely clear at this moment but it is fair to say that static/stabilized stripe order hinders superconductivity.

The significance of spin/charge ordered states in the physics of hole doped cuprates have been demonstrated by recent comparative study of the thermoelectric transport in Y123 and Eu-LSCO systems [23], which showed that the stripe order is the generic non-SC ground state of hole doped cuprates. This work [23] also suggests that the Fermi-surface reconstruction, as seen by quantum oscillation experiments [24], is possibly due to the breaking of the lattice translational symmetry by the incipient stripe ordering. The appearance of field induced charge ordering in Y123 [25] lends further support that stripe order is indeed a generic feature for all different families of high-$T_c$ cuprates. The study of the thermoelectric transport properties also showed that the basic features of the stripe related phase diagrams for Y123 and Eu-LSCO are essentially the same and the charged stripe phase exists over an extended doping range from $p = 0.08$ to 0.18. If the electronic ground state is indeed stripe ordered in this extended range of hole concentrations then any modification of the striped state by any means can possibly lead to a change in the superconducting state including the transition temperature itself. Based on this assumption we put forward a simple scenario as an explanation for the *p*-dependent anomalous IE observed in cuprates as follows. Isotopic substitution changes the lattice dynamics even if only by changing the characteristic phonon energies. In a situation where lattice and QP degrees of freedoms are strongly coupled, such a change in the phonon spectrum could also modify the stripe order and consequently change $T_c$. In fact direct experimental evidence in favor of such a proposal can be found in isotope effect experiments on Nd-LSCO [21]. The stronger the stripe correlations, the stronger would be the effect of isotope substitutions. It is worth noticing that this proposed change in the superconducting transition temperature is not directly linked to the change in the energy scale of the characteristic harmonic phonons as envisaged in the conventional BCS framework. Our scenario naturally explains the anomalous increase in the IE in the vicinity of the 1/8$^{th}$ doping. At this hole content stripes are pinned strongly by the lattice and the QP-lattice coupling is at its strongest. Therefore, change in the phonon spectrum affects the striped state most significantly around the 1/8$^{th}$ anomaly, thereby inducing a large shift in $T_c$ upon isotopic substitution. To illustrate the above points, we show the *p*-dependent oxygen ($O^{16} \rightarrow O^{18}$) IE for LSCO (details can be found in refs. [4, 10, 26]) together with the characteristic temperature, $T_{s0}$, taken from ref. [23], in Fig. 1. $T_{s0}$ is the temperature where the normal state Seebeck coefficient changes its sign due to a reconstruction of the Fermi-surface as stripe correlations set in and break the translational symmetry [23]. The correspondence between the doping evolutions of $T_{s0}(p)$ and $α(p)$ is striking. To our knowledge, Fig. 1 shows the first clear link between $α(p)$ with any other *p*-dependent parameter relevant to the physics of hole doped cuprates.

It is important to note that $α(p)$ remains small but finite outside the *p* range from 0.08 to 0.18. The small IE in these doping regions could be due to a truly phonon mediated pairing mechanism. In fact a relatively small electron-phonon (e-ph) coupling constant ~ 0.4 was found by Conte *et al.* [17], whereas the coupling strength due to spin fluctuations (e-sf) were found to be ~ 1.1. In the case of a joint pairing mechanism, where the phonons only play a minor role, a greatly reduced IE is expected. In its simplest form, the IE in the case of mixed mechanism, can be expressed as [5, 27], $\alpha = 0.5 [1 - \lambda_I/(\lambda_I + \lambda_{e-ph})]$, where $\lambda_I$ is the non-phononic (e.g., $\lambda_{e-sf}$) coupling constant and $\lambda_{e-ph}$ is the coupling constant due to the e-ph interaction. Using the values obtained by Conte *et al.* [17], the pairing related IE from phonons turns out to be ~ 0.13, a value not far away from the experimentally observed $\alpha(p)$ outside the *p*-range from 0.08 to 0.18. This in turn implies that pairing related IE in cuprates is masked by the strong stripe related effects over a significant doping range.

Quite interestingly, the proposed scenario also explains naturally the disorder induced enhancement in IE [4, 5] qualitatively. For example, it is believed that Zn pins stripe fluctuations [28, 29] and at the same time for a given *p*, IE increases in Zn substituted samples [4, 5, 10]. Here IE gets enhanced because the isotopic substitution had a greater effect in modifying the relatively more stabilized stripe ordered state in the Zn doped systems compared to the Zn-free ones. In the OD side Zn has a much smaller effect [4] because the stripe orders are severely weakened or even absent and Zn is unable to have a significant effect on stripe dynamics. As an exception to the rule, Zn substitution is seen to diminish the maximally enhanced $\alpha(p)$ at $p \sim 0.125$ [10]. At this precise doping the stripe order in LSCO is at its strongest and Zn plays no further appreciable role in pinning. Instead, the spin vacancies created by Zn hampers the integrity of the already stabilized stripe order [30, 31]. Ni substitution, on the other hand, has almost no effect on the IE in LSCO [32]. This would simply indicate that non-magnetic Zn and magnetic Ni play vastly different roles in stabilizing the stripe correlations in cuprates. This may also have a possible link with the interesting observation that unlike Ni, Zn suppresses $T_c$ much more effectively.

A question might be raised regarding the degree of validity of comparing $T_{s0}(p)$ of Y123 and Eu-LSCO with the $\alpha(p)$ for pure LSCO. To address this issue, we would like to mention the following facts. Nd/Eu is doped in LSCO only to amplify the effects of stripe correlations by stabilizing the order via anisotropic lattice distortion [22]. The proposed scenario predicts a further systematic enhancement in the $\alpha(p)$ in Nd/Eu-LSCO compared to that found in pure LSCO, as confirmed partly in ref. [21]. It is worth mentioning that the dynamic magnetic correlations in Nd doped and pure LSCO are essentially identical [22]. $T_{s0}(p)$ data for Y123 is shown in Fig. 1 to illustrate the possible universality of the proposed scenario. Y123 with its double $CuO_2$ planes, CuO chains, and much larger $T_c$, is a very different system structurally from LSCO. Nevertheless, the $T_{s0}(p)$ data for LSCO and Y123 are almost indistinguishable.

So far, systematic isotope effect experiments were mainly done on systems related to Y123 and LSCO. It has been shown in ref. [23] that the non-SC ground states of both Y123 and LSCO are indeed stripe ordered over an extended $p$-range. If it is true for all the holed doped high-$T_c$ cuprates, then we expect that the $α(p)$ trend should be very similar to those shown in Fig. 1 in other systems, even though a quantitative difference should exist as the degree of stripe correlations vary from system to system.

To summarize, we have put forward a scenario where the isotope effect in hole doped cuprates over an extended range of hole contents is dominated by the interplay between stripe and SC correlations. Our proposal provides with a way to reconcile the significant IE with non-phononic pairing in high-$T_c$ cuprates. It is up to the theorists now to develop the proper formalism, as the solution to the long standing puzzle regarding the *anomalous α(p)* and its relevance to Cooper pairing seems to be nearer.

The authors acknowledge the AS-ICTP, Trieste, Italy, for the hospitality.

**Figure caption**

Figure 1(color online): $T_{s0}(p)$ (taken from ref. [23]) for Eu-LSCO (blue squares) and Y123 (green diamonds), and the oxygen isotope exponent, $\alpha(p)$ (taken from refs. [4, 10, 26]) for pure LSCO (red circles). The full red curve is drawn as a guide to the eyes.

Figure 1

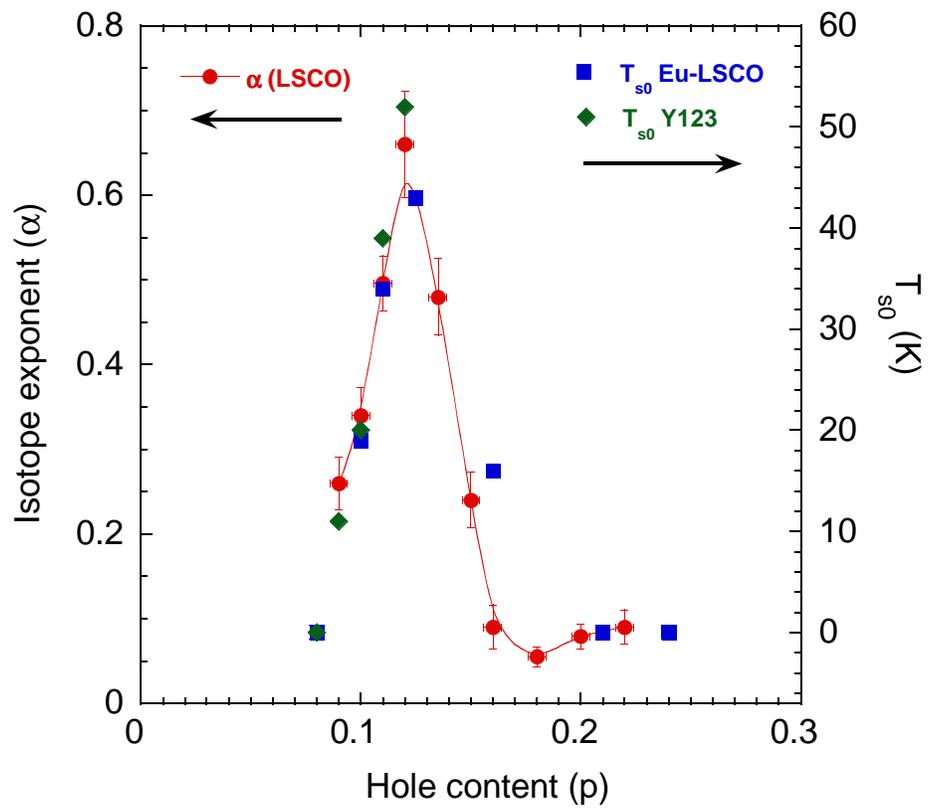